\documentclass{jpsj-suppl}
\usepackage{mathrsfs} 
\usepackage{amsmath,amssymb}

\def\Tr{\mathop{{\rm Tr}}}
\def\ket#1{|#1\rangle }
\def\bra#1{\langle #1|}
\def\braket#1#2{\langle #1|#2\rangle}

\def\section#1{}
\def\bms#1{{\boldsymbol{#1}}}
 
\title{Application of Uniform Matrix Product State to Quantum Phase Transition with a Periodicity Change}
\author{Hiroshi \textsc{Ueda}\thanks{E-mail address: ueda@aquarius.mp.es.osaka-u.ac.jp} and Isao \textsc{Maruyama}\thanks{E-mail address: maru@mp.es.osaka-u.ac.jp} 
}

\inst{Department of Material Engineering Science, Graduate School of Engineering Science, Osaka University, 
Toyonaka, Osaka 560-8531, Japan 
}
\abst{
As a method beyond the mean-field analysis,
a matrix product state (MPS) with incommensurate periodicity is applied to detect phase transitions accompanied with periodicity change, 
where the incommensurate MPS is generated by acting local-spin-rotation operators with the incommensurate periodicity on a uniform MPS. 
As a commensurate/commensurate change,
we calculate the partial ferro -- perfect ferro phase transition in the $S=1/2$ Heisenberg model
and its critical exponent of the magnetization curve.
As a commensurate/incommensurate change,
we calculate the $S=1$ Heisenberg model with bilinear and biquadratic interactions which has periodicity change in the spin-spin correlation function. 
}
\kword{
matrix product state, magnetization, phase transition, antiferromagnetic spin chain, variational method, incommensurate periodicity
}
\begin{document}
\maketitle
\paragraph{Introduction}
To detect the quantum phase transition and to characterize the ground state of the quantum many-body Hamiltonian are highly challenging tasks.
The mean-field analysis based on the one-body approximation is one of successful and useful methods
but completely neglects quantum entanglement between two particles, which is quite important to represent the spin singlet state.
The matrix product state (MPS)~\cite{Ostlund:PRL75, Rommer:PRB55} is one way to handle quantum entanglement gradually starting from the mean-field analysis.
Especially, in one-dimensional (1D) quantum spin systems, many numerical studies by use of the density matrix renormalization group (DMRG) method~\cite{White:PRL69-PRB48, Peschel:Springer, RevModPhys.77.259} reveal that the MPS becomes a good variational state of finitely correlated systems. 
An MPS in spin $S=1/2$ $L$-site systems is represented as
$\ket{\Psi_{\rm v}} = \sum_{\bms{\sigma}}^{} \Tr \big[ \prod_{j=1}^{L} A^{\sigma_j}_{j} \big] \ket{{\bms{\sigma}}}, \label{vertex}$
where $\sigma_j = \uparrow, \downarrow$, $\bms{\sigma} = \{\sigma_1 \cdots \sigma_L\}$ and $A^{\sigma_j}_{j}$ is an $m \times m$ matrix at $j$-th site. 
This MPS is called the vertex-type MPS.~\cite{Baxter:book, Sierra1997505} 
One of the variants of the MPS is called the interaction-round-a-face-type (IRF-type) MPS~\cite{Baxter:book, Sierra1997505}: 
$\ket{\Psi_{\rm i}} = \sum_{\bms{\sigma}}^{} \Tr \big[ \prod_{j=1}^{L} A^{\sigma_j, \sigma_{j+1}}_{j} \big] \ket{\bms{\sigma}}, \label{IRF}$
where $\sigma_{L+1} = \sigma_1$ and $A^{\sigma_j, \sigma_{j+1}}_{j}$ is also an $m \times m$ matrix at $j$-th site. 
The vertex-type MPS is a special case of the IRF-type MPS, namely $A^{\sigma_j, \sigma_{j+1}}_{j} = A^{\sigma_j}_{j}$.
Hereafter, we will focus on the IRF-type MPS.
The MPS can express a translation-invariant state in the thermodynamic limit ($L\rightarrow \infty$) 
with finite degrees of freedom using a uniform matrix $A^{\sigma_j, \sigma_{j+1}}_{j} = A^{\sigma_j, \sigma_{j+1}}_{}$,
in spite of the infinite dimension of the Hilbert space in the thermodynamic limit.
The infinite time-evolving block decimation (iTEBD)~\cite{PhysRevLett.98.070201} uses this advantage,
where the Suzuki-Trotter decomposition is used for optimization of the MPS
and breaks the translation-invariance of the MPS.~\cite{PhysRevLett.98.070201, Chen:PRB82}.
In order to discuss the translational period precisely, the Suzuki-Trotter decomposition is not used in this work.

Even if the translation-invariance of the exact ground state in a spatially-homogeneous Hamiltonian is broken spontaneously, 
the uniform MPS may keep translation-invariance in the limit $m\rightarrow \infty$
using the linear combination of the degenerated ground states,
as checked for an anti-ferromagnetic Ising model~\cite{JPSJ.80.023001}.
However, for fixed and finite $m$,
the optimum variational state to represent $q$-site periodic state
is the $q$-site modulated MPS with $A^{\sigma_j, \sigma_{j+1}}_{j} = A^{\sigma_j, \sigma_{j+1}}_{{\rm mod}[j,q]}$
~\cite{JPSJ.80.023001},
where the commensurate period $q$ is a positive integer and the MPS with the $q=1$ period means the uniform, i.e.,  translation-invariant, MPS.
Since the degrees of freedom of the $q$-site modulated MPS is $q$-times larger than the uniform MPS,
ground states with long period $q$ require large computational memory to be represented by MPSs.
In this meaning, incommensurate period is impossible to be represented by the finite degrees of freedom.

To overcome this problem, we have introduced the MPS with incommensurate period by acting local-spin-rotation operators recently~\cite{Ueda:unpublished}: 
$\ket{\Psi_{\rm i}(Q, {\bf n})} = \big[ \prod_{j}^{} \exp(-i Q j \hat{{\bf S}}_j \cdot {\bf n}) \big] \ket{\Psi_{\rm i}},\label{incomme}$
where a parameter $Q$ and a unit vector ${\bf n}$ mean a pitch of the rotation angle and a rotational axis in the spin space, respectively. 
The operator $\hat{{\bf S}}_j$ represents the spin operator at $j$-th site. 
Note that a uniform MPS $\ket{\Psi_{\rm i}}$ is a special case of the incommensurate MPS (IC-MPS) $\ket{\Psi_{\rm i}(Q, {\bf n})}$, namely 
$\ket{\Psi_{\rm i}} = \ket{\Psi_{\rm i}(0, {\bf n})}$. 
We have already confirmed that this IC-MPS can detect the incommensurate properties in spin-spin correlation functions in $S=1/2$ zigzag chain.~\cite{Ueda:unpublished}
However, it is not well understood how the transition point and the critical phenomena, for example critical exponents, are captured by the IC-MPS. 

In this work, we apply the IRF-type IC-MPS to following two problems about the critical phenomena at the phase boundary.
One is (i) the commensurate/commensurate change at the phase transition between partial and perfect ferro-magnetism 
in an $S=1/2$ antiferromagnetic (AF) Heisenberg chain. 
The other is (ii) the commensurate/IC (C/IC) change in an $S=1$ spin system with bilinear and biquadratic interactions. 
The mean-field analysis must be reproduced by our method in the case of $m=1$ and vertex-type IC-MPS. 
The rigorous result will be obtained in the limit $m\rightarrow \infty$.
Our motivation is to interpolate the two limits.

\paragraph{Optimization Condition}
Let us show details of our numerical calculations before the discussion of our results. 
In this work, we use the IRF-type MPS $\ket{\Psi_{\rm i}(Q, {\bf n})}$ with $A^{\sigma_j, \sigma_{j+1}}_{j} = A^{\sigma_j, \sigma_{j+1}}_{}$. 
The $m\times m$ complex matrix-elements, pitch $Q$, and axis ${\bf n}$ are optimized so that the variational energy 
$\bra{\Psi_{\rm i}(Q, {\bf n})} \hat{H} \ket{\Psi_{\rm i}(Q, {\bf n})} / \braket{\Psi_{\rm i}(Q, {\bf n})}{\Psi_{\rm i}(Q, {\bf n})}$
for the Hamiltonian $\hat{H}$ becomes minimum by using the modified Powell method~\cite{Numerical_Recipes:book}.
Additionally, 10--2000 initial different matrices are prepared and optimized in each Hamiltonian parameter to avoid obtaining a false variational energy. 
The rotational axis ${\bf n}$ is taken as $(0,0,1)$ in following results. 

\paragraph{Commensurate/Commensurate Change}
First, we evaluate a critical exponent of the magnetization near the phase transition 
between partial ferro -- perfect ferro in the $S=1/2$ AF Heisenberg chain under the uniform longitudinal magnetic field:  
$\hat{H}_1 = \sum_{j} ( \hat{{\bf s}}_{j} \cdot \hat{{\bf s}}_{j+1} - h^z \hat{s}^z_j ), \label{ham_hb}$
where $\hat{{\bf s}}_{j}$ and $h^z$ mean the spin $S=1/2$ operator at $j$-th site and the longitudinal magnetic field, respectively. 
The rigorous result~\cite{Takahashi} shows that the transition point is at $h^z_{\rm c} = 2$ with the exponent $\beta = 1/2$,
while the mean-field calculation shows $\beta = 1$.

\begin{figure}[Htb]
\begin{minipage}[t]{0.45\linewidth}
   \centering
   \resizebox{7cm}{!}{\includegraphics{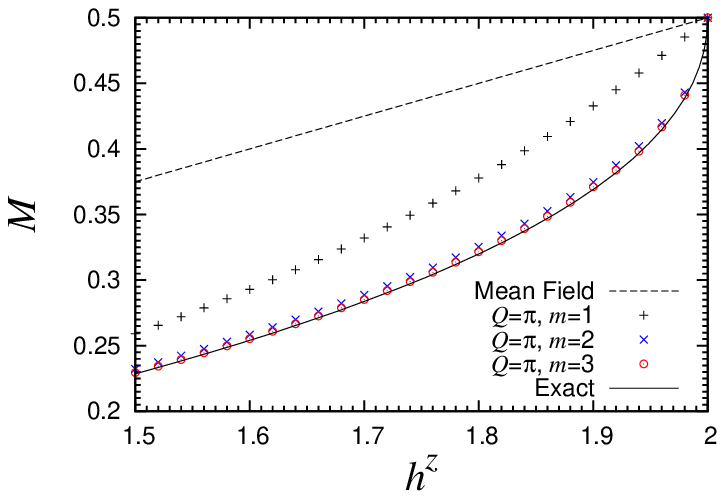}}\\
   \caption{(Color online) Magnetization curve as a function of $h^z$ for $\hat{H}_1$.}
   \label{fig:mag}
\end{minipage}
\hspace{0.1\linewidth}
\begin{minipage}[t]{0.45\linewidth}
  \centering
  \resizebox{7cm}{!}{\includegraphics{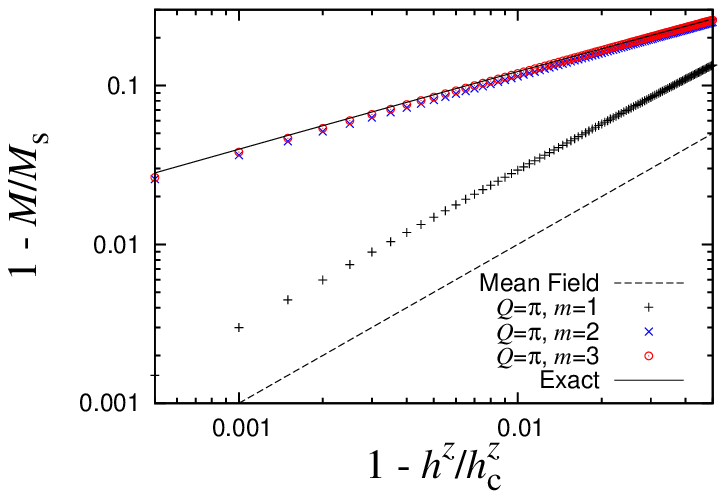}}\\
  \caption{(Color online) Asymptotic behavior of magnetization curve approaching the saturated magnetization $M_{\rm s}$ with the double logarithmic plot for $\hat{H}_1$.}
  \label{fig:ce}
\end{minipage}
\end{figure}

As a numerical result,
the magnetization curve of the $S=1/2$ Heisenberg chain by using the IC-MPS is shown in Fig.~\ref{fig:mag}. 
The magnetization per site is represented by $M$. 
The pitch $Q$ becomes $\pi$ after the optimization irrespective of the magnetic field $h^z$.~\cite{Ueda:unpublished} 
The broken line is calculated by the mean-filed analysis, which is $M = h^z/4$.
This mean-filed solution is also obtained if we use the vertex-type IC-MPS with $m=1$.
In contrast, the IRF-type IC-MPS with $m=1$ in this figure is better than the mean-field clearly, 
because the IRF-type MPS is able to generate non-zero entanglement entropy even if $m=1$.
However, the IC-MPS with $m=1$ overestimates the magnetization in all partial-ferro region. 
The overestimation is remarkably suppressed with increasing $m$ as shown in Fig.~\ref{fig:mag}. 
In the case of $m=3$, the estimation error in the magnetization becomes less than one percent. 

We evaluate the critical exponent $\beta$ using the following relation: $1 - M/M_{\rm s} =  (1 - h^z/h^z_c)^\beta. $
Figure~\ref{fig:ce} shows the asymptotic behavior of the magnetization approaching the saturated magnetization $M_{\rm s}=1/2$.
The exponent $\beta$ evaluated by the IC-MPS with $m=1$ agrees with that from the mean-field analysis, namely $\beta=1$, within error of five percent. 
The exponents $\beta$ evaluated by the IC-MPS with $m=2$ and $3$ consist with that from the rigorous analysis, namely $\beta=1/2$, within error of five percent for the region depicted in Fig.~\ref{fig:ce}. 
It can be concluded that the IC-MPS with $m=2$ is sufficiently accurate for the present model
while the IC-MPS with $m=1$ shows similar behavior to the mean-field result.

The previous study on the Ising chain with a uniform transverse magnetic field~\cite{Chen:PRB82} suggests 
that the vertex-type MPS leads to mean-filed behavior for any finite $m$ in the vicinity of the transition point 
because the MPS can represent only finitely correlated state with a finite $m$. 
However, we cannot find such a crossover behavior for the IC-MPS with $m=2$ of the present model 
in the asymptotic behavior of $1-M/M_{\rm s}$ down to $1-h^z/h^z_c \sim 1\times10^{-5}$.  

\paragraph{Commensurate/Incommensurate Change}
To apply our method to the C/IC change,
we consider the $S=1$ spin chain with bilinear and biquadratic interactions: 
$\hat{H}_2 = \sum_{j} \cos \theta \hat{{\bf S}}_{j} \cdot \hat{{\bf S}}_{j+1} + \sin \theta (\hat{{\bf S}}_{j} \cdot \hat{{\bf S}}_{j+1})^2, \label{ham2-4}$
where $\hat{ {\bf S} }_j$ is the spin $S=1$ operator at $j$-th site, and $\theta$ determines the ratio of intensity between bilinear and biquadratic interactions. 
Previous numerical works reported that the IC pitch of spin-spin correlation functions approaches a commensurate pitch with increasing $\theta$ up to 
$\theta_{\rm c} = \pi/4$.~\cite{PhysRevB.53.3304, PhysRevB.74.144426}
This point of $\theta_{\rm c}$ is also a gapfull--gapless phase transition point, and $k=\pm2\pi/3$ spin quadrupolar correlations are dominant in $\theta=[\pi/4,\pi/2]$.

\begin{figure}[Htb]
  \centering
  \resizebox{7cm}{!}{\includegraphics{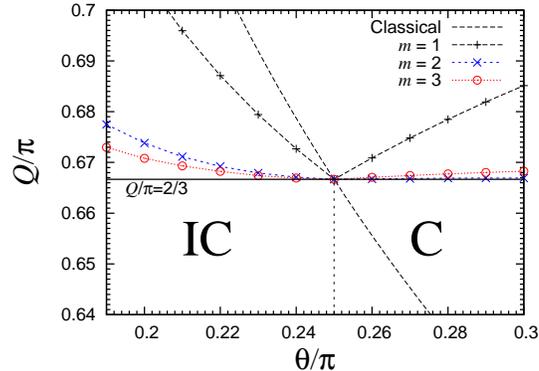}}\\
  \caption{(Color online) Incommensurate pitch $Q$ as a function of the model parameter $\theta$ for $\hat{H}_2$. 
Characters C and IC are abbreviations of commensurate and incommensurate, respectively.}
  \label{fig:ic-c}
\end{figure}

As a numerical result, the change of periodicity $Q$ as a function of the model parameter $\theta$ is shown in Fig.~\ref{fig:ic-c}. 
The broken line represents the pitch obtained from the classical vector spin model, which is $2 \cos Q = - {\rm cotan} \: \theta$. 
The commensurate-correlation region cannot be detected by the classical vector analysis. 
On the other hand, in the analysis with the IC-MPS, the C/IC change is revealed immediately with increasing $m$. 
The pitch $Q/\pi$ in the commensurate region agrees with $2/3$ within error of 0.5 \% for $m=3$.
Thus, the pitch $Q$ detects the property change of the ground-state's correlations in this model. 
It can be concluded that our method  detects the C/IC change with increasing $m$. 

\paragraph{Summary}
We apply a variational method with the IRF-type IC-MPS to phase transitions in (i) the $S=1/2$ AF Heisenberg chain under magnetic field and 
(ii) the $S=1$ AF Heisenberg chain with bilinear and biquadratic interactions. 
As the result, our method has successfully identified the exact phase transition points for both models.
Since the MPS is identical to the mean-field analysis in the simple limit and will become the exact ground state in the limit $m\rightarrow \infty$,
this method has interpolated the two limits successfully as shown in Fig.~\ref{fig:mag}. 
Especially, we find that the exact critical exponent $\beta=1/2$ is obtained with small $m$.
This becomes a useful property of this method. 
Moreover,
in the bilinear-biquadratic Hamiltonian, 
we have succeeded to detect the C/IC change by small $m$,
while the analysis of the classical vector spin model fails to detect it.
The pitch $Q$ of the commensurate region is immediately converged with increase of $m$ to $Q=2\pi/3$, which is consistent with the period of the dominant spin-spin correlations in the previous studies. 
In the incommensurate region, the convergence of $Q$ with respect to $m$ is not enough. 
The improvement of optimization method to treat large $m$ is one of the remaining issues. 
In addition,
application to the transverse filed Ising model 
is one of the future issues
in order to compare the previous study.~\cite{Chen:PRB82} 

\paragraph{Acknowledgment}
This work was supported in part by a Grant-in-Aid for JSPS Fellows and Grant-in-Aid No. 20740214, Global COE Program (Core Research and Engineering of Advanced Materials-Interdisciplinary Education Center for Materials Science) from the Ministry of Education, Culture, Sports, Science and Technology of Japan. 


\end{document}